\documentclass[aps,pra,preprint,showpacs,preprintnumbers,amsmath,amssymb]{revtex4-1}
\usepackage[utf8x]{inputenc}
\usepackage{graphicx}
\usepackage{amsmath}
\usepackage[all]{xy}
\usepackage[spanish, english]{babel}
\usepackage{setspace}
\usepackage{latexsym}
\usepackage{float}
\usepackage{subcaption}
\usepackage[figurename=Figure]{caption}

\usepackage{txfonts}
\usepackage{amssymb}
\usepackage{graphicx}
\title{Light Spring compression in a multi frequency Raman amplifier}

\begin{document}

\title{ Light Spring amplification   in a multi-frequency Raman amplifier }
\author{J. A. Arteaga$^1$, A. Serbeto$^1$, K. H. Tsui$^1$, J. T. Mendon\c{c}a$^2$ }
\affiliation{$^1$Instituto de F\'isica, Universidade Federal Fluminense, Campus da Praia Vermelha, Niter\'oi, RJ,  24210-346, Brasil,}
\email{johny@if.uff.br}
\affiliation{$^2$IPFN, Instituto Superior T\'ecnico, Universidade de Lisboa, 1049-001 Lisboa, Portugal}

\begin{abstract}
We propose to amplify and compress an ultrashort Light Spring laser seed  with a long Gaussian-shaped laser pump through Raman amplification. This  Light Spring, which  has  a helical spatio-temporal intensity profile,  can be built on the superposition of three distinct laser frequency components.  In order to get  an independent frequency  amplification, two criteria are established. Besides these criteria, a non equal frequency separation is necessary  to avoid resonance overlapping when three or more frequencies are involved.  The independent set of equations, which describes the wave-wave interaction in a plasma, is solved numerically for two different Light Spring configurations. In both cases, the amplification and transversal compression of the seed laser pulse have been observed, with a final profile similar to that of the usual Gaussian-shaped seed pulses. In  addition, two different kinds of  helical plasma waves are excited.  
\end{abstract}
\maketitle
\section{Introduction}
 Since the idea proposed by Malkin et al. \cite{malkin} to improve the actual laser intensity by compression mechanism of the scattered laser beam in  plasma through the stimulated Raman backscattering, many techniques have been explored to mitigate many unwanted effects. Such techniques include chirping the laser seed frequency \cite{shirping}, imposing laser coherence conditions after the amplification \cite{coherence}, and controlling of the laser parameters to extend the Backward Raman Amplification  beyond the  transversal relativistic instability \cite{relativistica},  where a Gaussian-shape laser seed is amplified and compressed in a final horseshoe-shape \cite{trines}. On the other hand, Barth  and Fisch \cite{nathaniel} have introduced the concept of multi-frequency Raman amplifier to obtain a high efficiency Raman amplification using a laser seed wave packet with two separated frequencies, in a similar way to single-frequency seed pulse. In this case two criteria are established in order to avoid resonance overlap.   

The advent of a new class of lasers with a donut-shaped intensity profile with a helical phase  gives rise to many applications in different branches of laser science. These kinds of laser beam have been conceived by Allen et al. \cite{allen} in such a way that are well described by a  family of Laguerre-Gaussian (LG) modes, which have the capability to carry and transfer the Orbital Angular Momentum (OAM) to matter \cite{yao}. Specifically, in a Raman amplifier, Viera et al. \cite{titonature} using a Particle-in-cell (PIC) simulation describe  various scenarios of OAM transfer to a Langmuir plasma wave using a LG pump and a seed laser,  where the total OAM gained by the plasma wave corresponds to the difference of OAM among the pump and seed lasers, i.e.,  $\ell_w \hbar- \ell_s \hbar = \ell_p \hbar$ \cite{titoprl}.

In order to get closer to the new generation of ultra high laser intensity, it is mandatory that the laser seed pulse, in a Raman amplifier, should have an ultrashort temporal duration, of the order of femtoseconds or attoseconds, leading to a broad frequency spectrum inside the wave packet. In this sense, Pariente and Quéré \cite{pariente} have introduced the idea of linearly correlated frequencies inside the pulse with the modal index $\ell$ of the superposed LG modes, which gives a  spatio-temporal correlated Light Spring (LS). This LS beam has the special feature that both intensity and phase get the same helical structure. These have recently been used to  excite a twisted plasma wake-field in laser-plasma interaction studies \cite{jviera}, capable of accelerating helical electron beams.     

In this paper, the concept of using an ultrashort LS laser as a seed interacting  with a Gaussian pump laser in a cold plasma through the Raman backscattering mechanism is introduced. This will lead to the amplification and compression of the LS signal.  

\section{Multi-Frequency Raman Amplifier}
Basically, a  single frequency Raman backward amplifier in a cold and under-dense plasma can be expressed in the paraxial approximation  by the  following set of  three-wave coupled equations:
\begin{eqnarray}
\left( -\frac{ic^2}{2\omega_1} \nabla_\perp^2 +\frac{\partial}{\partial t} + c\frac{\partial}{\partial z} \right)\,a_{1} & =& \sqrt{\frac{\omega_1 \omega_p}{2}} \hat{f}_{1} \, b_{1}   \\
\left( -\frac{ic^2}{2\omega_1}  \nabla_\perp^2 +\frac{\partial}{\partial t} - c\frac{\partial}{\partial z} \right)\,b_{1} & =& - \sqrt{\frac{\omega_1 \omega_p}{2}} \, \hat{f}^*_{1} \, a_{1} \\
\left( -\frac{iS_e^2}{2\omega_p} \nabla_\perp^2 + \frac{\partial}{\partial t} - \frac{ k_{f_1}}{ \omega_{f_1}}S_e^2\frac{\partial}{\partial z}  \right)  \hat{f}_{1} &=& - \sqrt{\frac{\omega_1 \omega_p}{2}} \, a_{1} \, b_1^*,
\end{eqnarray}
where $a_1, b_1$ are the normalized vector potentials for a circular polarized electromagnetic  pump wave propagating along z-axis and a counter-propagating electromagnetic seed, respectively; while $\hat{f}_1 = -i c k_{f1} (\omega_1/2\omega_p^2)^{1/2} \, \delta n_1/n_0$ is  the normalized electrostatic field  of the Langmuir plasma wave, where $k_{f1}$ is the longitudinal plasma wavenumber, $\omega_p = (4\pi n_0 e^2/ m)^{1/2} $ is the plasma frequency, $n_0$ is the equilibrium plasma density, $\delta n_1$ is the perturbed plasma density and $S_e = \sqrt{3T_e/m_e}$ is the electron thermal velocity. Here, the Raman backscattering matching conditions are $\omega_{a_1} = \omega_{b_1} + \omega_{f_1}$ and $k_{a_1} = k_{b_1} + k_{f_1}$, whose the dispersion relations are  $\omega_{a_1,b_1} = \sqrt{c^2 k^2_{a_1,b_1} + \omega_p^2 }$ and $\omega_{f_1} = \sqrt{\omega_p^2 + S_e^2 k^2_{f_1}} $. In  Eqs (1)-(3) the under-dense cold plasma approximation have been used leading to  $\omega_1 \equiv \omega_{a_1}  \approx \omega_{b_1} \approx c k_{b_1} $, $\omega_{f_1} \approx \omega_p$, and $k_{f_1}= k_{a_1} + k_{b_1} \approx 2 \omega_1 / c $.

According to Ref. \cite{nathaniel}, in a multi-frequency Raman amplifier, two conditions are imposed in order to avoid overlap and get an independent amplification. The first one is that the nearest separation frequency  has to be  greater than the maximum bandwidth gain acquired by each mode, in the linear Raman stage (not pump depletion), that is, $\Delta  = \omega_j  - \omega_i > \textnormal{max}\{a_{j} \sqrt{\omega_{j} \omega_p/2}\}$ \cite{linearR}.  The second condition is that the 
 pulse duration,$\tau_{FWHM}$, should be greater than the  period of the resulting beat wave, that is, $\tau_{FWHM} > 2\pi/\Omega_{ji}= 4\pi /(\omega_j - \omega_i)$. Here, we will consider three  frequency components to describe the pump and seed wave packet, such that the frequency separation  is given by
\begin{eqnarray}
\omega_2 - \omega_1 = \Delta, \,\,\,\,\,\, \omega_3 -\omega_2 = \omega_2 -\omega_1 + \delta, \label{conditions}
\end{eqnarray} 
where the meaning of $\delta$ is explained a posteriori. \\
The total normalized  potential vectors of the pump and seed are described by
 \begin{equation}
 \vec {a} = \frac{1}{\sqrt{2}} \operatorname{Re} \left( \sum^{3}_{j=1}  a_j\, e^{i(\omega_j t  -k_j z)} \,\hat{e} \right), \, \, \, \, \, \vec {b} = \frac{1}{\sqrt{2}} \operatorname{Re} \left( \sum^{3}_{j=1} b_j\, e^{i(\omega_j t   + k_j z)} \,\hat{e} \right)
\end{equation}  
where  $\hat{e} = (\hat{x} + i\hat{y})/2$  is the circular polarization vector. The matching conditions are given by
\begin{eqnarray}
\omega_{a_j} = \omega_{b_j} + \omega_{f_j},  \,\,\,\, k_{f_j} = k_{a_j} + k_{b_j}, \,\,\,\, \textnormal{for}\, j=1,2,3,
\end{eqnarray}
with $\omega_{a_j,b_j} = \sqrt{c^2 k^2_{a_j,b_j} + \omega_p^2 }$ and $\omega_{f_j} = \sqrt{\omega_p^2 + S_e^2 k^2_{f_j}}. $
The  under-dense cold plasma approximation leads to   $\omega_{a_j,b_j} \approx c k_{a_j,b_j}$, making the down-shifted frequency negligible, i.e, $\omega_{b_j} \approx \omega_{a_j} \equiv \omega_j $. Hence, the set of equations to describe  three independent Raman amplifiers is given by 
\begin{eqnarray}
\hat{D}_{P_j} \,a_{j} & =& \sqrt{\frac{\omega_j}{\omega_1}} \, \hat{f}_j \, b_j ,  \label{ai}\\
\hat{D}_{S_j} \, b_{j} & =&  \sqrt{\frac{\omega_j}{\omega_1}} \,  \hat{f}^*_{j} \, a_{j}, \label{bi} \\
\hat{D}_{L_j}\, \hat{f}_{j} &= &- \frac{\omega_p}{\omega_{f_j}}\sqrt{\frac{\omega_j}{\omega_1}} \, a_j \, b^*_j, \label{fi} \,\,\,\,\,\,\,\,\, \textnormal{for}\, j=1,2,3,
\end{eqnarray}
where $\hat{D}_{P,S,L}$ are the differential operators associated to the pump, seed  and Langmuir waves defined as follows
\begin{eqnarray}
\hat{D}_{P,S_j} &=&  -\frac{i}{\sqrt{8}} \frac{\sqrt{\omega_1\omega_p}}{\omega_j} \bar{\nabla}_\perp^2 +\frac{\partial}{\partial t'} + \hat{k}_{p,s}\frac{\partial}{\partial z'}  ,\\
\hat{D}_{L_j}&= & -\frac{i }{\sqrt{8}} \frac{S^2_e}{c^2}  \frac{\sqrt{\omega_1\omega_p}}{\omega_{f_j}}\bar{ \nabla}_\perp^2  + \frac{\partial}{\partial t'} - \frac{k_{f_j}}{\omega_{f_j}} \frac{S_e^2}{c}\frac{\partial}{\partial z'},
\end{eqnarray}
with $\hat{k}_p = +1$ and $\hat{k}_s = -1$. 
Here, the time and space variables are normalized with $\sqrt{\omega_1 \omega_p / 2}$ and $\sqrt{\omega_1 \omega_p / 2}/c$, respectively. We should point out that in  Eqs (7-9)  the coupling  between the envelopes is only for the same  $j$-mode. In this set of equations the non-resonant terms, which correspond to coupling with  different $j$-modes,  are neglected since other Raman decay processes are avoided,  that is, $\omega_j \neq \omega_i + \omega_p$ for $i \neq j$, leading to  $\omega_p \neq \{\Delta, \Delta +\delta, 2\Delta + \delta\} $. As we can see,  on the right-hand-side (RHS)  of Eqs (\ref{ai} - \ref{fi}) we have complex exponential terms whose arguments are given by  $k_{a_i} +  k_{b_j} - k_{f_k} \approx k_i + k_j - 2k_k$, for $i \neq j \neq k$. Then, in order to avoid cross resonance, these arguments can not be null. This condition gives us the necessary requirement of non equal separation among the frequencies $\omega_1$ and $\omega_3$, respect to the central frequency, $\omega_2$, which corresponds to taking a small frequency shifting,  $\delta $,  as given by  Eq. (\ref{conditions}). For $\delta \ne 0$, it gives us
 fast phase terms, such as exp$\,(i(k_3  + k_1 - 2k_2)\,z$) $\approx$ exp$\,(2 i \delta z/c)$ and others, which  on average, do not contribute to the Raman resonant amplification. 

\section{Light Spring}
Following  Ref. \cite{pariente}, a LS beam can be generated  when  each  frequency inside the laser pulse is correlated with the LG  modal index $\ell$, 
 according to  \begin{equation}
 \ell(\omega) = \ell_0 + \frac{\Delta \ell }{\Delta \omega} (\omega-\omega_0),
\end{equation}
where $\omega_0$ is the central frequency corresponding to a modal index $\ell_0$,   $\Delta \omega$ is the spectral width, and $\Delta \ell$ is the variation of the modal index through $\Delta \omega$. The correlation given by Eq. (12) results in a  beat wave with a helical intensity profile, whose temporal pitch is given by $\tau_p = 2\pi \Delta \ell / \Delta \omega$. Considering two kinds of LS beam, which are uniquely differentiated by the slope, $\Delta \ell/ \Delta \omega$, as detailed in Fig. 1, the superposition of the three modes is represented as
\begin{equation}
\mathit{b} = \sum_{j=1}^{3} B_j (r, t)\,  e^{\mathit{  \varphi}_j (r,\theta, z)} , 
\end{equation}
where
\begin{equation}
B_j = B_{0j} (z,t) w_{0j} \sqrt{\frac{2p_j!}{\pi w_j^2 (p_j + |\ell_j|)!}}\left( \frac{\sqrt{2}r}{w_j} \right)^{|\ell_j|}L^{\ell_j} _p \left(\frac{2r^2}{w_j^2} \right) e^{-r^2/ w^2_j},
\end{equation}
is the LG mode amplitude
 and 
 \begin{equation}
 \varphi_j = i \left[ \frac{R_j z r^2}{z^2_{0j} + R_j^2} \,-\,(2p_j +|\ell_j| +1)\arctan(z/z_{0j})\,+\,   \ell_j \phi \right], \label{phase}
\end{equation}
is the respective screw phase. Here $B_{0j}$ represents the amplitude of each mode and $w_{0j} = z_{0j}(1 + R_j^2 z^2 /z_{0j}^2)$ is the normalized transverse  waist, which is a function of the normalized Rayleigh length, $z_{0j} = w_{0j}^2$. In this equation, for $\ell= 0$, we have the Gouy phase, which gives the curvature evolution of the wavefront, with $R_j = \sqrt{\omega_1\omega_p}/\omega_j$.  The last term in $\varphi_j$ generates the helical structure of  the LG mode, with $\ell$ being the modal index, which gives us the amount of OAM carried by the mode.\\
\begin{figure}[h!]
\centering  \includegraphics[width=3.5in]{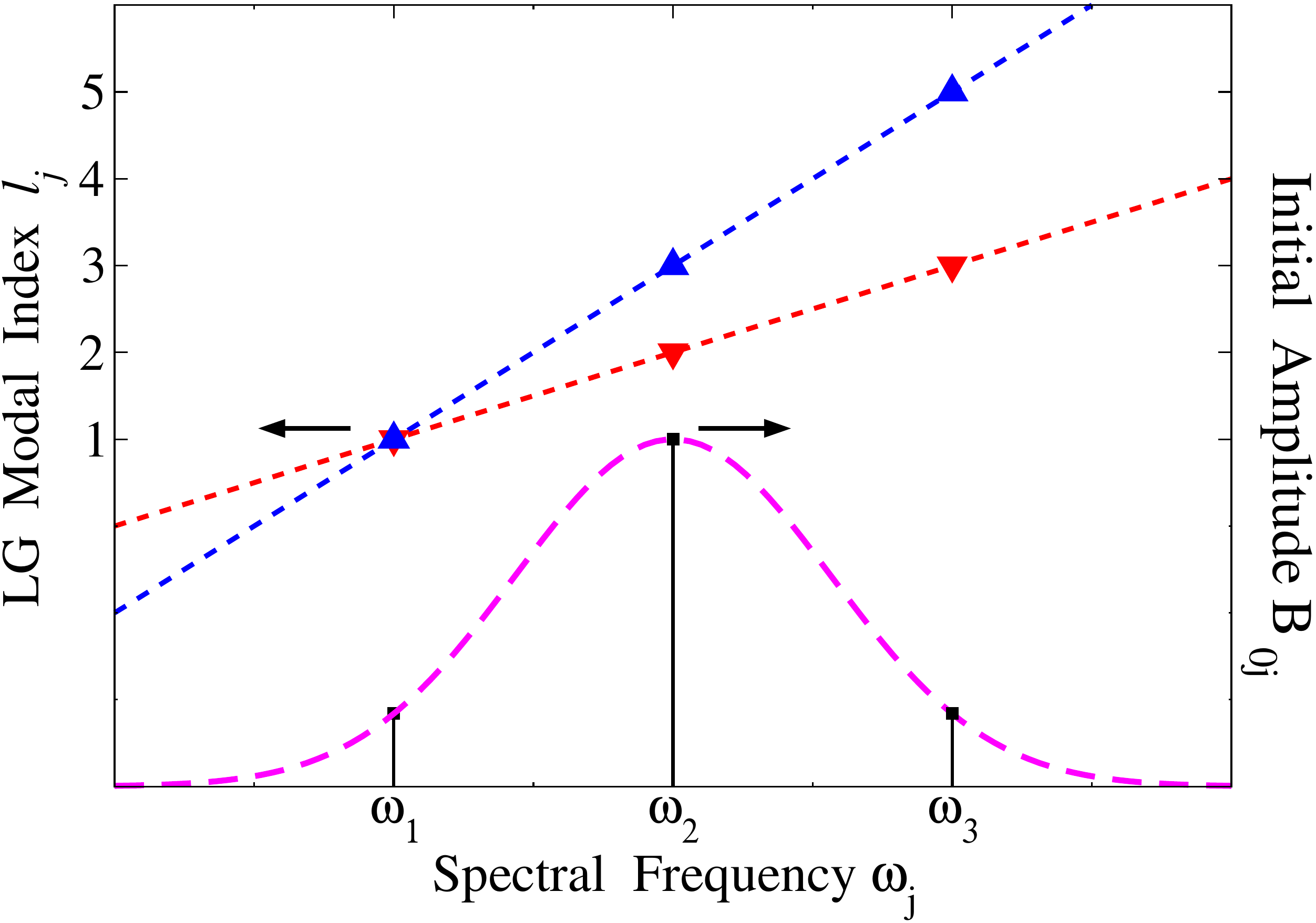}
\caption{\small{Two modal correlation indexes with different slopes for a seed pulse composed by three frequencies (pink dotted line). The first one has a slope $\Delta \ell / \Delta \omega = 2$ (blue dotted line) with integer  modal indexes $\{1,3,5\}$. The second case,  $\Delta \ell / \Delta \omega = 1$ (red dotted line) with integer modal indexes $\{1,2,3\}$. In both cases, each mode has an initial  Gaussian profile, where the modes with  frequencies $\omega_1$ and $\omega_3$, respectively, have an amplitude corresponding to $20\%$ of the central mode $\omega_2$. }  }
\label{correlation}
\normalsize
\end{figure}
This set of coupled equations (\ref{ai}-\ref{fi}) is independently  solved for each  $j$-mode, for the two cases mentioned in Fig. (1). Hence, the final  beat wave amplitude will be  given by the  superposition of each  envelope   amplitude, i.e,  for the beat wave seed  we have $\vec{b} = \sum_{j} \vec{b}_j$. For simplicity  we will consider an initial transverse waist   $w_j = w_{0,j}= w_0$ large enough to avoid diffraction effects on the seed laser pulse  during the interaction time. Also, the LG modes will be considered to have only one ring in its intensity profile, in such a way that $p_j = 0$.  The pump laser pulse will be assumed to have a Gaussian transversal structure with the same waist of the seed pulse and have a temporal duration of picoseconds, in order to have the Raman process reaching the nonlinear stage. Here we will consider a plasma with an electron temperature $T_e = 5$eV,  electron density $n = 5\times 10^{18}$ cm$^{-3}$ ($\omega_p \approx 126$ THz),  central frequency $\omega_2 = 2\pi\times375$ THz ($\lambda \sim 0.8\mu$m),  low frequency $\omega_1 = 2\pi\times345$ THz and  high frequency $\omega_3 =  (2\pi\times405 + 1)$ THz. With these frequencies,  we have the frequency shift, $\Delta  = 2\pi \times30 $ THz, and a small mismatching parameter, $\delta = 1$ THz. The initial central seed mode is assumed to have an amplitude $B_{02} = 2\times10^{-2}$,  while the initial pump amplitude is set at $a_j = a_0 = 0.1$, leading to $\textnormal{max}\{4 a_{j} \sqrt{\omega_{j} \omega_p/2}\} = 4 a_0 \sqrt{\omega_3\omega_p/2} \approx 1.6 \times10^{14}< \Delta$, which satisfies the first criteria for a multi-frequency Raman interaction. For the second criteria, each mode that composes the pulse has a temporal duration  $\tau_{FWHM} = 84 fs > 2\pi\Omega_{ij}= 33 fs $, which is also satisfied. 
\begin{figure}[b!]	
	\begin{subfigure}[b]{3.2in}
		\centering
		\includegraphics[width=3.2in]{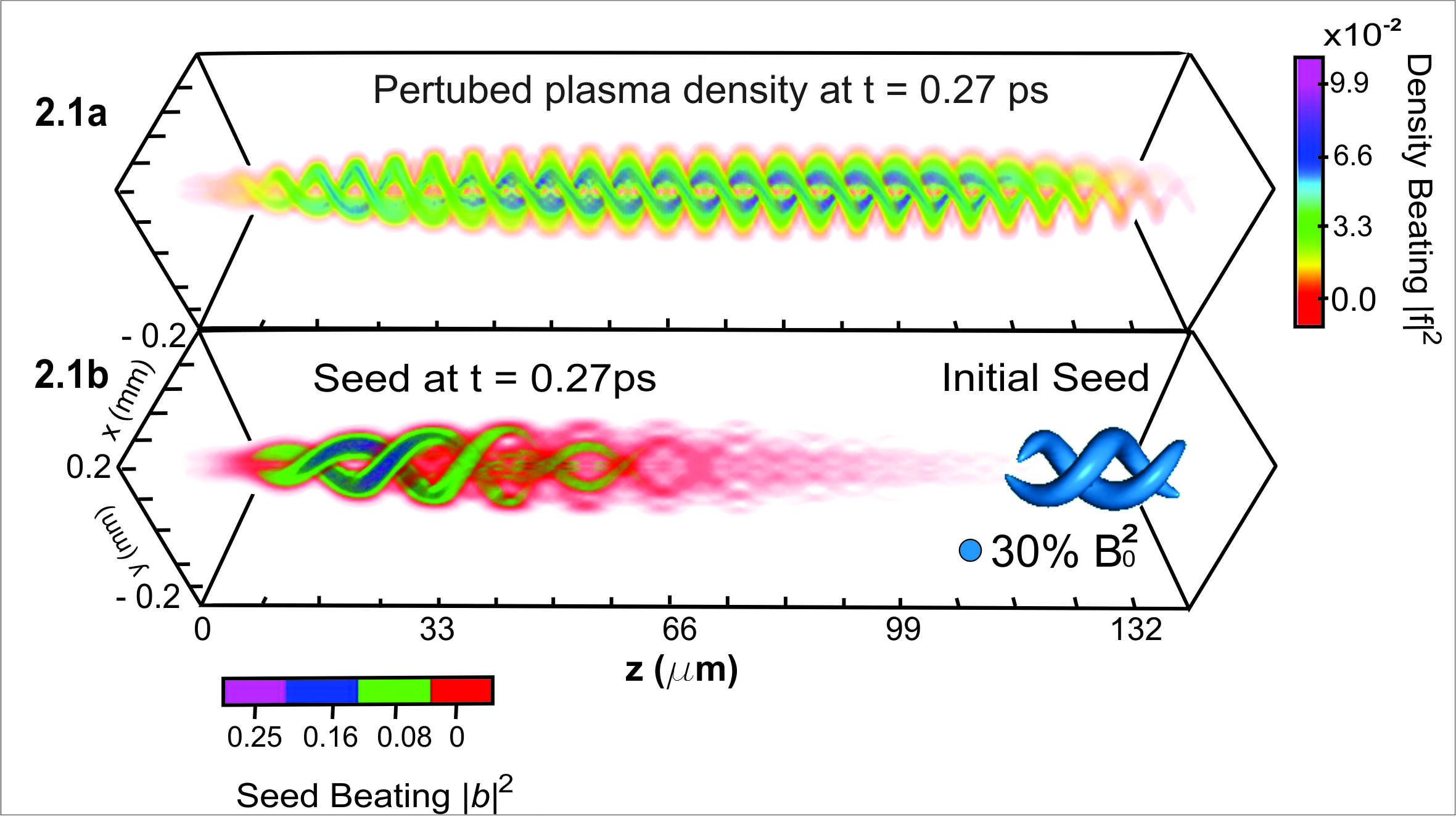}
		\caption{{ 2.1.}$\,\,$ Slope $\Delta\ell/\Delta\omega = 2 $ }
	\end{subfigure}
    \quad
	\begin{subfigure}[b]{3in}
		\centering
		\includegraphics[width=3in]{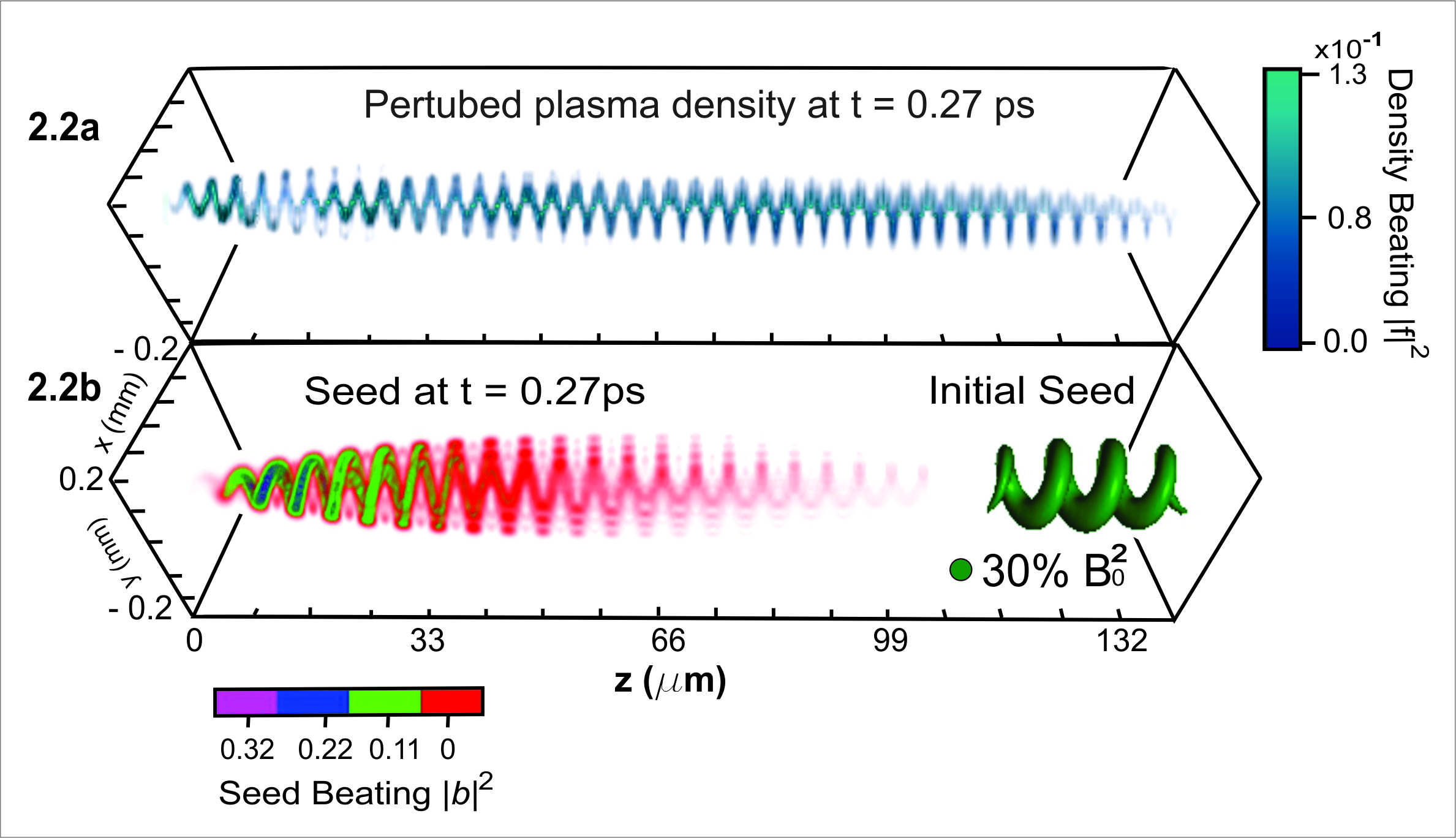}
		\caption{{ 2.2.}$\,\,$ Slope $\Delta\ell/\Delta\omega = 1 $ }\label{fig:2a}		
	\end{subfigure}
	\caption{\small{Figures 2.1a and 2.2a   show the final intensity profile of the  perturbed plasma wave. Figures 2.1b and 2.2b, at right side, show the isosurface corresponding to $30\%$ of the initial peak laser seed intensity, which final intensity  profile is presented at left side.  }}\label{fig:2}
\end{figure}
On the right side of Figures 2.1b and 2.2b, the initial LS seeds are presented for the two cases of slope respectively mentioned in Figure 1. Each slope produces one kind of LS as is reported in \cite{pariente}, where for the case $\Delta \ell/\Delta \omega = 1$ (Fig. 2.2b) the seed LS presents only one coil, while for  $\Delta \ell/\Delta \omega = 2$ (Fig 2.1b) two intertwined coils are generated. These LS structures on the intensity profile are described by the beating of the three near frequency modes as follows
\begin{equation}
\vec{b} \cdot \vec{b}^* = |b|^2 = \sum_{j=1}^3 |b_j|^2 + 2 \textnormal{Re}\, \bigg\{\sum_{\substack{i,j\\ i \neq j,\, i\,>j}}^3 b_i b_j^*\, e^{i[(k_j - k_i)z -(\omega_j -\omega_i)t]} \bigg\} \label{beating}
\end{equation}
In both cases, during the non linear stage, the LS seed is amplified and compressed in the  front of the pulse    
 as is represented in Figs. 2.1b and 2.2b. Fig. 3.1  shows the iso-surface (orange), at left of the box,  taken at 25$\%$ from the final maximum value of the seed, and at right of the box, the  iso-surface (blue) taken   at 25$\%$ of the seed  initial maximum value. Similarly,  Fig. 3.2 represents the final (purple) and initial (green) iso-surfaces  for  one coil seed. Inside the gray surface in Figs 3.1 and 3.2 is  the leading pulse of the seed that looks very similar to the horseshoe shape of the conventional Gaussian-shape pulses   \cite{titonature, trines} because of the  transverse compression in its front. Outside the gray surface we can observe the formation of a second spring that  keeps the same feature of the leading spring  but with less intensity corresponding to the sub-pulses generated by the energy exchange during Raman interaction. Figs. 2.1a and 2.2a show the beat wave generated by the three perturbed modes of the plasma density, which are computed in the same way as in Eq.(\ref{beating}).  In both cases, the perturbed plasma density acquires a helical structure according to the kind of the LS seed, obeying the individual OAM conservation \cite{titoprl}, and keeping the same properties of the conventional one frequency Raman amplifier, that is, the density maximum is located in the same region of maximum seed beating. 
\begin{figure}[t!]	
	\begin{subfigure}[t]{3.1in}
		\centering
		\includegraphics[width=3.1in]{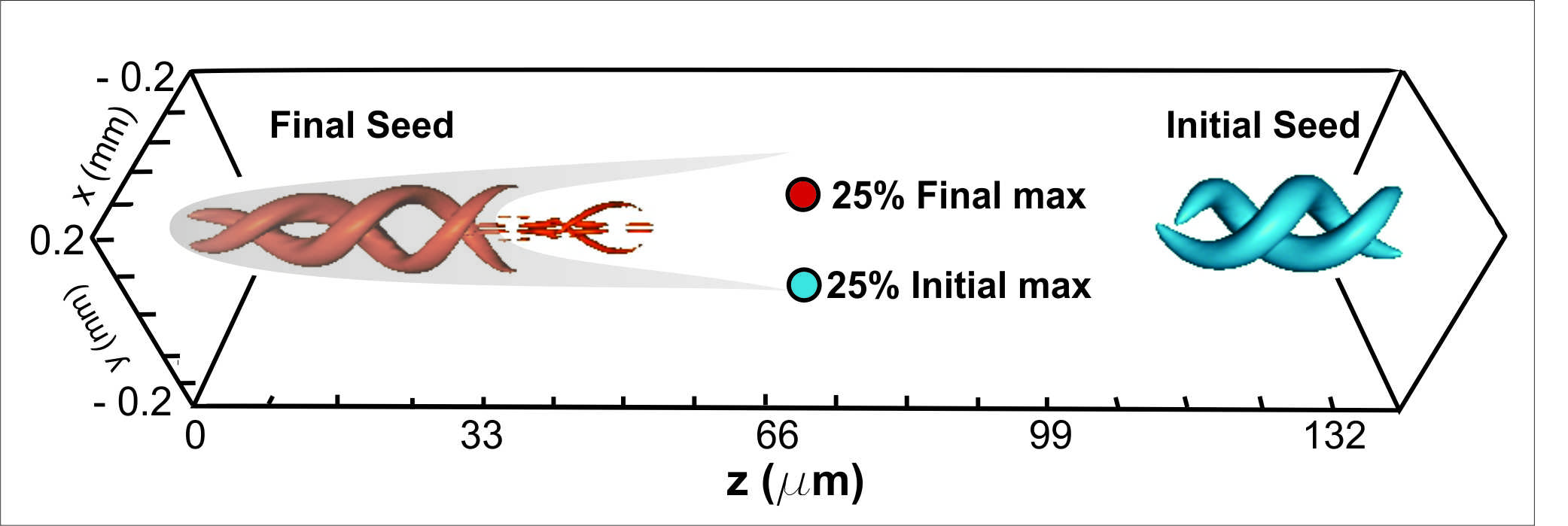}
		\caption{{ 3.1.}$\,\,$ Slope $\Delta\ell/\Delta\omega = 2 $ }
	\end{subfigure}
    \quad
	\begin{subfigure}[t]{3.1in}
		\centering
		\includegraphics[width=3.1in]{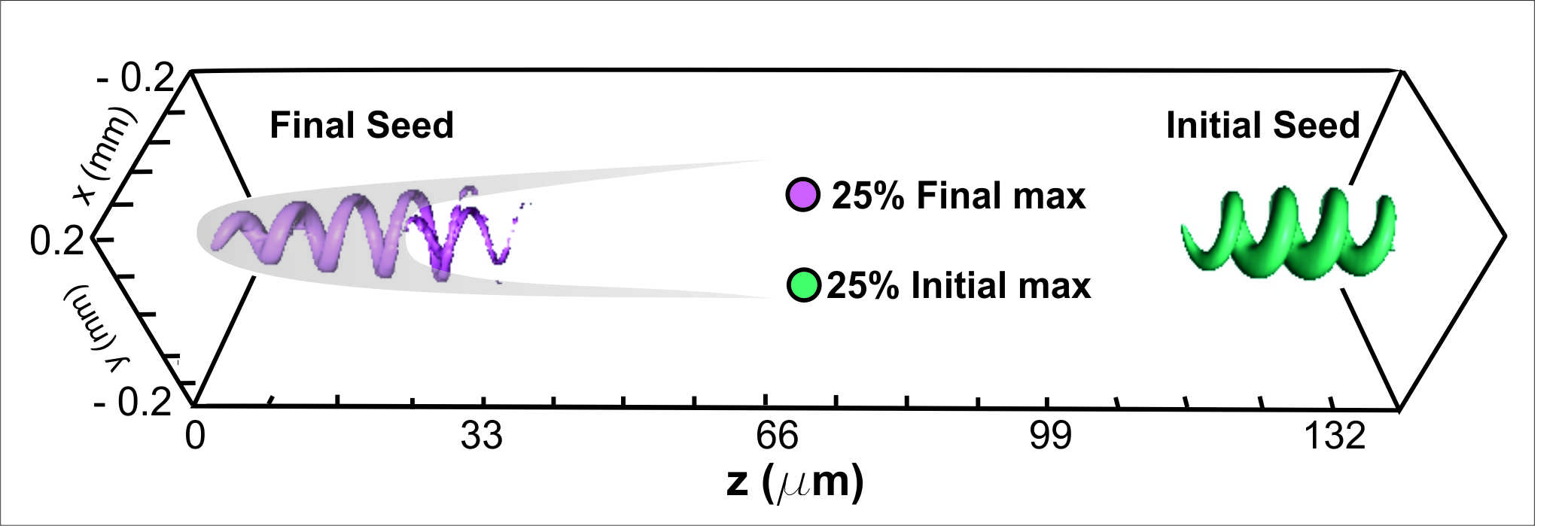}
		\caption{{ 3.2.}$\,\,$ Slope $\Delta\ell/\Delta\omega = 1 $ }\label{fig:2a}		
	\end{subfigure}
	\caption{\small{ Seed iso-surfaces for the initial and final stages for the cases in Fig 2.1b and Fig 2.2b. For both cases we have a iso-surface taken at the $25\%$ of the initial and final maximum value.  The horse-shoe shape of the main pulse is represented by the gray surface in the final stage of both cases.   }}\label{fig:2}
\end{figure}
\section{Conclusions}
In conclusion, we have implemented a LS seed pulse through the correlation of the spectral width with the modal index from the LG modes. It is verified that, depending on the slope $\Delta\ell/\Delta\omega$, various intertwined coils can be generated by the spring. It is important to note that, as represented by the blue and red dotted lines in Fig 1, along the spectral frequency of the pulse there exists a continuum of frequencies that is correlated according to Eq.(12), which leads to a large number of non integer modal index that  are not solutions of the paraxial equation. These non integer modes cause  the diffraction on the edge of LS, while keeping  its helical structure, as discussed in Refs. \cite{pariente, berry,spin}. With respect to the dynamics of the LS seeds, we employ a triple Raman amplifier where the separation between each  mode has to satisfy some minimum requirements to get its independence. These requirements were proposed and  tested by analyzing the Fourier spectrum in a 1D PIC simulation for a  single, double and triple Raman amplifier \cite{nathaniel}. Also, in a multi-frequency  Raman scheme, when the down shifted frequency is neglected, i.e $\omega_{a_j} = \omega_{b_j}$,  the   separation  between a neighboring pair of frequencies has to be	different from  another pair as given by the $\delta$ factor defined in Eq.(4), which supports  the individual evolution of our triple wave interaction  model. From the well known shape  of a single-frequency  Raman amplifier, the transversal  compression of the initial LS laser seed gives  rise to what we call  a {\it horseshoe Light Spring}. However, when we compare the thickness of the final  and initial iso-surfaces in Figs. 3.1 and 3.2, these are essentially  the same, opposing to  the already known  longitudinal compression in the non linear stage  of a single-frequency Raman \cite{malkin}. It may be presumably related  to the broadening suffered by  each mode in the linear Raman stage, given by $a_j \sqrt{\omega_j \omega_p}$, that depends on the mode frequency $\omega_j$, leading to  different compression rates and  causing the maximum of the lower frequencies slipping ahead of the highest one. Also, the transverse coupling  between the Gaussian-shape pump with each of  the seed mode is different because of the vortex size is proportional to the amount of OAM carried by each mode.  
In addition, we have shown the possibility of having a spiral phase in a plasma wave and a helical profile in its intensity. This opens the possibility of applying this technique to generate  helical electron beams  \cite{hemsingnature, johny}. Finally, we are conscious of the limitations in the three wave interaction scheme discussed here.  In order to predict with more accuracy the dynamics of the process, it is mandatory to use 3D PIC simulations to check if the longitudinal compression occurs,  and to explore  the effect of  instabilities on the LS beam in a multi frequency Raman scheme.

\section*{Acknowledgment}
We thank Kenan Qu, Qing Jia and  Nathaniel Fisch, from University of Princeton, for their fruitful comments to develop this work, and Martin Silva for graphic support.
This research is supported by the National Council of Research 
and Technological Development (CNPQ) of Brazil


\begin{thebibliography}{99}

\bibitem{malkin}
V. M. Malkin, G. Shvets, and N. J. Fisch {\sl Phys. Rev. Lett.}, {82}, 4448 (1999)

\bibitem{shirping}
Z. Toroker, V. Malkin, and N. J. Fisch, {\sl Phys. Rev. Lett.}, {109}, 085003 (2012)

\bibitem{coherence} 
M. R. Edwards,  K. Qu, J. M. Mikhailova,  and N. J. Fisch, {\sl Physics of Plasmas}, {24}, 103110 (2017) 

\bibitem{relativistica}
V. M. Malkin, Z. Toroker, and N. J. Fisch {\sl Phys. Rev. E.}, 
 {90}, 063110 (2014)

 \bibitem{trines}
 R. M. G. M. Trines, F. Fiúza, R. Bingham, R. A. Fonseca, L. O. Silva, R. A. Cairns and P. A. Norreys {\sl Nature Physics}, {7}, 87 (2011)
 
\bibitem{nathaniel}
I. Barth and N. J. Fisch, {\sl Phys. Rev. E}, {97}, 033201 (2018)
\bibitem{allen}
L. Allen, M. W. Beijersbergen, R.J.C Spreeuw, J. P. Woerdman  {\sl Phys. Rev. A}, {\bf 45}, 8185 (1992). 
 
 
\bibitem{yao}
A. Yao and M. Padgett, {\sl Adv. Opt. Photon.}, {\bf 3}, 161 (2011)

 
 \bibitem{titonature}
J. Vieira, R. M. G. M. Trines, E. P. Alves, R. A. Fonseca, J. T. Mendonça, R. Bingham, P. Norreys, L. O. Silva {\sl Nature Communications}, {\bf 7}, 10371 (2016)

\bibitem{titoprl}
J. T. Mendonça, B. Thidé, H. Then , {\sl Phys. Rev. Letters}, {\bf 102}, 185005 (2009)

\bibitem{pariente}
G. Pariente and F. Quéré {\sl Optics Letters }, {40}, 9 (2015)

\bibitem{spin}
J. T. Mendonça, A. Serbeto and J. Vieira, {\sl Sc. Rep.}, {8}, 7817 (2018).

\bibitem{jviera}
J. Vieira, J. T. Mendonça, and F. Quéré, {\sl Phys. Rev. Lett.},  {121}, 054801 (2018) 

\bibitem{linearR}
V. M. Malkin, G. Shvets, and N. J. Fisch, {\sl Physics of Plasmas}, {7}, 2232, (2000) 

\bibitem{berry}
M. V. Berry {\sl J. Opt. A: Pure Appl. Opt.}, {6}, 259 (2004)

\bibitem{hemsingnature}
E. Hemsing, A. Knyazik, M. Dunning, D. Xiang, A. Marinelli, C. Hast, J.  B. Rosenzweig {\sl Nature Physics}, {\bf 9}, 549 (2013).

\bibitem{johny}
J. A. Arteaga, A. Serbeto, J. T. Mendonça, K. H. Tsui, and L. F. Monteiro {\sl Physics of Plasmas}, {24}, 123108 (2017)




\end{thebibliography}
\end{document}